# Acoustic Hologram Optimisation Using Automatic Differentiation

Tatsuki Fushimi[1, 2, 5] *, Kenta Yamamoto[1, 3, 5], Yoichi Ochiai[1,2,4]

[1] R&D Center for Digital Nature, University of Tsukuba, Tsukuba 305-8550, Japan

[2] Faculty of Library, Information and Media Science, University of Tsukuba, Tsukuba 305-8550, Japan

[3] Graduate School of Library, Information and Media Studies, University of Tsukuba, Tsukuba 305-8550, Japan

[4] Pixie Dust Technologies, Inc., Tokyo 101-0061, Japan

[5] These authors contributed equally: Tatsuki Fushimi, Kenta Yamamoto.

* Corresponding author

Acoustic holograms are the keystone of modern acoustics. It encodes three-dimensional acoustic fields in two dimensions, and its quality determine the performance of acoustic systems. Optimisation methods that control only the phase of an acoustic wave are considered inferior to methods that control both the amplitude and phase of the wave. In this paper, we present Diff-PAT, an acoustic hologram optimisation algorithm with automatic differentiation. We demonstrate that our method achieves superior accuracy than conventional methods. The performance of Diff-PAT was evaluated by randomly generating 1000 sets of up to 32 control points for single-sided arrays and single-axis arrays. The improved acoustic hologram can be used in wide range of applications of PATs without introducing any changes to existing systems that control the PATs. In addition, we applied Diff-PAT to acoustic metamaterial and achieved an >8 dB increase in the peak noise-to-signal ratio of acoustic hologram.

# Introduction

Acoustic hologram is becoming an imperative part of a wide range of acoustics applications such as in the fields of medicine[1–3], biology[4–8], and engineering[9–11]. The reconstruction accuracy of the acoustic field from the hologram plays a significant role in determining the performance of the system. Thus, developing an acoustic hologram optimiser that can generate holograms with accurate field reconstruction is of significant interest of the field. Recent advancement of airborne phased array transducers (PATs) have enabled new applications such as airborne ultrasound tactile display (AUTD)[12,13] and acoustic levitation[14–17]. Acoustic levitation in particular is used for digital fabrication[18,19], sample holding in medicine[20,21], physics[22,23] and chemistry[24] and display applications[25–30]. In all of these applications, acoustic holograms must generate a pressure control/focal point at a specified position. The control point is modulated at low frequencies to create haptic sensations that are sensible by human hands in an AUTD[12,13]. Marzo et al. demonstrated that the control points can be converted into levitation points by adding a twin-trap hologram in acoustic levitation[31,32].

Generating an acoustic hologram for a single control point in space using PATs is trivial; however, it has been a significant challenge to generate a hologram that can create more than one control point in space. Multiple control points are becoming a necessary part of PAT applications, and low-quality acoustic hologram can lead to inaccurate haptic sensation or levitation points. To address this challenge, Long et al. proposed Eigensolver and Tikhonov based regularisation in 2014[13]. Marzo & Drinkwater proposed the iterative back propagation (IBP) method in 2018[32], and Plasencia et al. proposed GS-PAT in 2020[33]. GS-PAT and IBP are both modified versions of the Gerchberg-Saxton algorithm[34]. Most recently, Sakiyama et al.[35] demonstrated acoustic hologram optimisation with the Levenberg-Marquardt algorithm (LMA); they examined the accuracy of ultrasonic stimulation in real and simulated environment. LMA and IBP optimise only the phase of the transducer array, and GS-PAT and Eigensolver can optimise both the amplitude and phase of the acoustic hologram. Placensia et al.[33] demonstrated that hologram optimisation methods with amplitude control (i.e. GS-PAT and Eigensolver) achieve higher-quality holograms than phase-only methods such as IBP, and that the Eigensolver method achieves the best optimised result because it seeks the global solution.

In this paper, we propose Diff-PAT, a phase-only, gradient-descent algorithm with automatic differentiation. We demonstrate that Diff-PAT has superior accuracy than the conventional Eigensolver and GS-PAT with both amplitude and phase control. Automatic differentiation is different from other differentiation operations (such as numerical differentiation and symbolic differentiation). This method calculates a derivative of a function at high accuracy by applying chain rules to each elementary operation of a given function. It is commonly used in machine learning. Our work was inspired by Peng et al.[36], who demonstrated this automatic differentiation based method for optical holograms. In comparison to acoustic holograms, optical holograms have been studied for a longer period, and a number of hologram optimisation methods have been developed. However, Peng et al. demonstrated that the method using automatic differentiation outperforms conventional methods such as the Gerchberg-Saxton method and the analytical differentiation method (Wirtinger Holography[37]). The challenge in optimising an optical or acoustic holograms is based on the fact that the loss function converts a complex number to a real number (non-holomorphic objective[37]). Whilst this function is challenging to differentiate analytically (the derivative of the function is not just a constant[37]), latest automatic differentiation packages[38–40] can differentiate these functions with high precision[41] and ease[36]. Achieving a high-quality acoustic hologram without amplitude control in PATs represents a significant contribution to the literature, as it allows PAT controllers to retain a simple design. Furthermore, we demonstrated the versatility of Diff-PAT by employing it for the optimisation of acoustic metamaterial (phase plate[42–45]). With regard to phase plates, the iterative angular spectrum approach (IASA) developed by Melde et al.[42] is the standard method for optimizing acoustic holograms for arbitrary acoustic fields. We achieved enhanced performance by simply applying Diff-PAT to the optimisation process, and without altering the fundamental design of Melde et al.

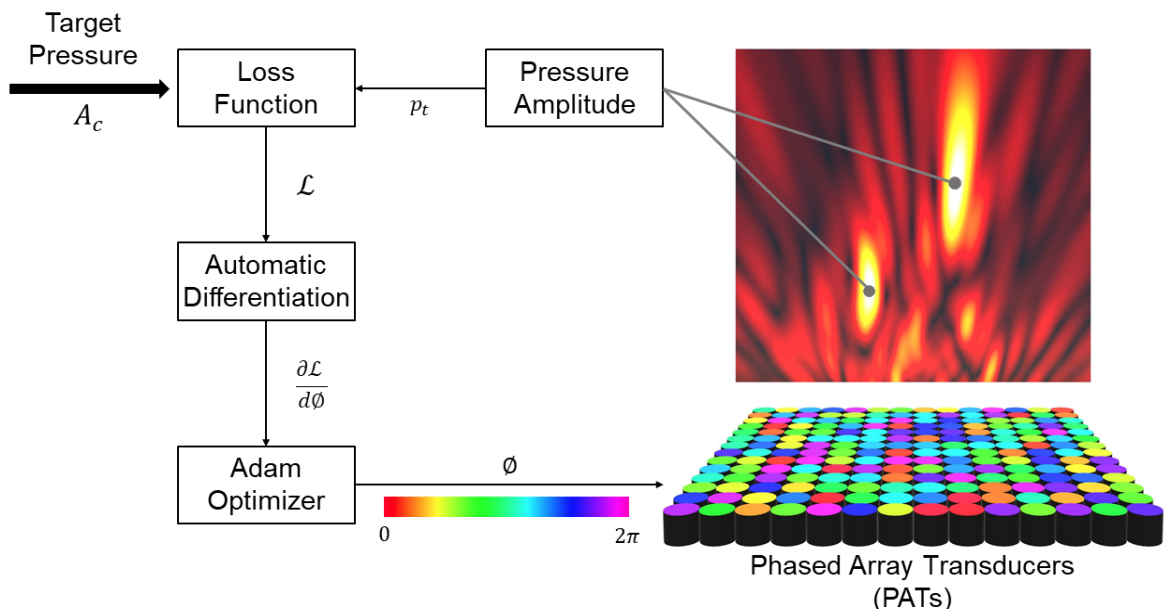

**Fig. 1 System overview for Diff-PAT.** Loss function is evaluated by comparing the target and current acoustic amplitude, and automatic differentiation is used to calculate the derivative of the loss function.

The system overview of Diff-PAT is as shown in Fig. 1. We used JAX (ver. 0.2.5) on Python 3.6.9 to perform automatic differentiation and optimisation[40]. To identify a suitable acoustic hologram ($\phi_n$) for a given control point $x_c$, target amplitude $A_c$, and transducer position $x_t$, we adopted the Adam optimiser[46], which enables efficient stochastic optimisation with only first-order gradients. Before the optimisation of hologram $\phi_n$, we defined the following optimisation problem and objective function $\mathcal{L}$:

$$\underset{\phi_n}{minimize} \ \mathcal{L}(\phi_n, x_c, A_c, x_t)$$

where $\mathcal{L}(\phi_n, x_c, A_c, x_t) = \sum_{c=0}^{C}(A_c - |p_t(\phi_n, x_c, x_t)|)^2$. $C$ is the total number of the control points, and $p_t(\phi_n, x_c, x_t)$ is the total acoustic pressure at $x_c$ (see "Methods" for details on $p_t$). In order to evaluate the effectiveness of Diff-PAT in PATs applications, acoustic hologram was generated for three setups of arrays (see Fig. 2a for layout); (1) a single-sided square array of 14×14 transducers (M = 196), (2) two square arrays of 16×16 transducers (M = 512), which face each other with a separation distance of 0.2355 m (i.e. a single-axis configuration) and (3) a single-sided square array of 32×32 transducers (M=1024). The single-sided array is a common arrangement for AUTDs, and the single-axis array is a popular configuration for acoustic levitation[14,47–49]. The single-sided square array with M = 1024 was included in the evaluation in order to allow for the increasing demand of 1000+ transducer PATs[18,50–52]. The initial guess of the phase was randomly generated for all transducers. The gradient is calculated by reverse-mode automatic differentiation according to the computational graph of the objective function $\mathcal{L}$, and the phase is optimised by updating itself with the Adam optimiser on JAX. The required hyperparameters for the Adam optimiser, $\alpha$, $\beta_1$, $\beta_2$, and $\varepsilon$ were set as; 0.1, 0.9, 0.999, and $1 \times 10^{-8}$, respectively.

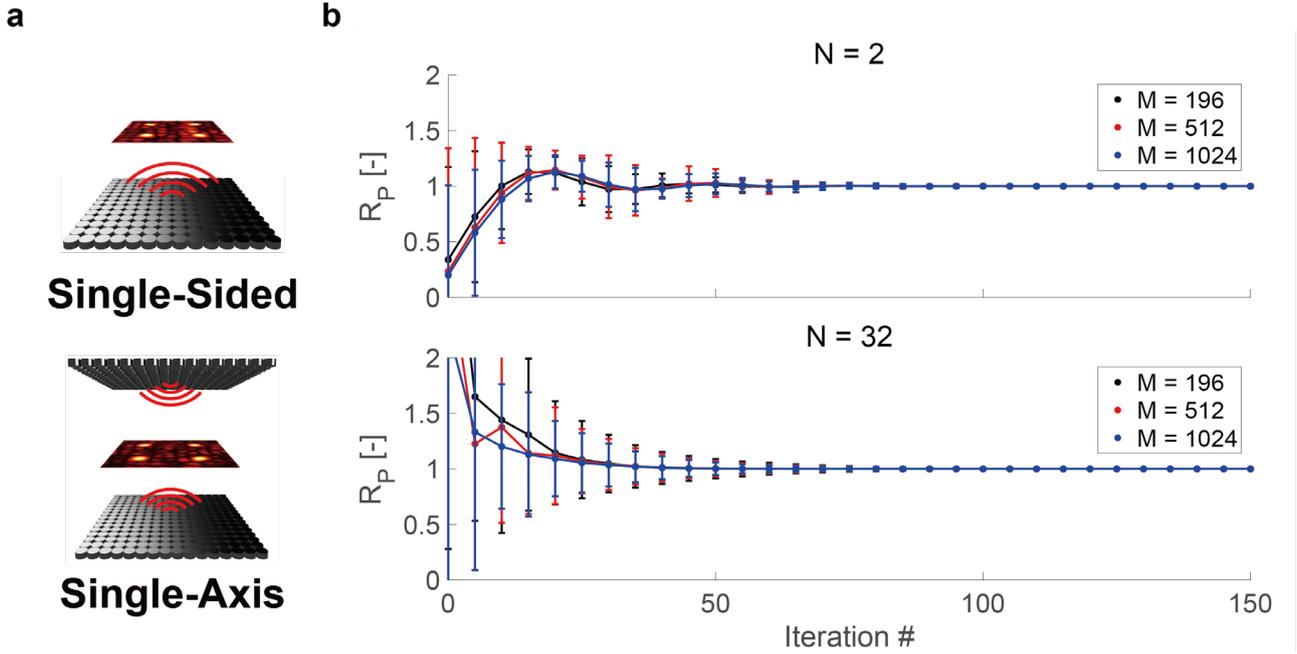

**Fig. 2 Array configuration and the convergence plot.** (**a**) Configuration of single-sided and single-axis PATs. (**b**) Convergence of optimisation function $\mathcal{L}$ by the iteration number, control point number (N) and transducer number (M). The convergence is evaluated by $R_p$. The data point and the error bar show the mean and standard deviation of $R_p$, respectively, for the population of randomised 1000 sets of control points.

## Results and Discussions

### Convergence rate of Diff-PAT

In the following section, the performance of Diff-PAT is evaluated. Because the algorithm must endure a wide range of combinations of control point positions and amplitudes, the control points were randomly generated, as described in the Methods section.

We show that the optimisation function converges quickly to the target value (evaluated by the ratio $R_p = \frac{|p_t(x_c)|}{A_c}$ between the target and current acoustic pressure amplitude), as shown in Fig. 2b (see Data

Availability for results at each iteration). Increasing the number of control points (N) has a negative effect on the convergence rate of the solution, and the length of the error bar (which shows the standard deviation of $R_p$) increases. However, the increase in the number of transducers has a negligible effect on the convergence rate (Fig. 2b). We confirm that our algorithm converges sufficiently between 100-150 iterations in Fig. 2b, and the number of iterations was set to 150 in all evaluations of Diff-PAT in this study. The convergence rate of the algorithm can be further improved by hyperparameter tuning; however, this is beyond the scope of the current study.

**Accuracy of Diff-PAT and comparison to conventional solvers**

The performance of Diff-PAT was compared with that of conventional algorithms in PATs which were made available by the authors of GS-PAT in C++[33]. Here, we compare Diff-PAT with the Eigensolver, corrected Eigensolver, and GS-PAT. The results are shown in the box-and-whisker plot in Fig. 3. Each algorithm was tasked to optimise the acoustic hologram for the same dataset, which included 1000 sets of randomised control points and amplitude (GS-PAT's iteration number was set to 100, as in their study[33]). Raw data for Fig. 3 can be accessed as stated in the Data Availability section.

Interestingly, when the control number is small (N = 2), Eigensolver performs poorly (median values for M = 196, 512 and 1024 are 0.737, 0.728 and 0.721, respectively, in Fig. 3). This large deviation is considered to be caused by the tendency of the regularisation policy to "homogenise transducer's output rather than reconstruction accuracy"[33]. This performance can be improved by using the corrected Eigensolver method,

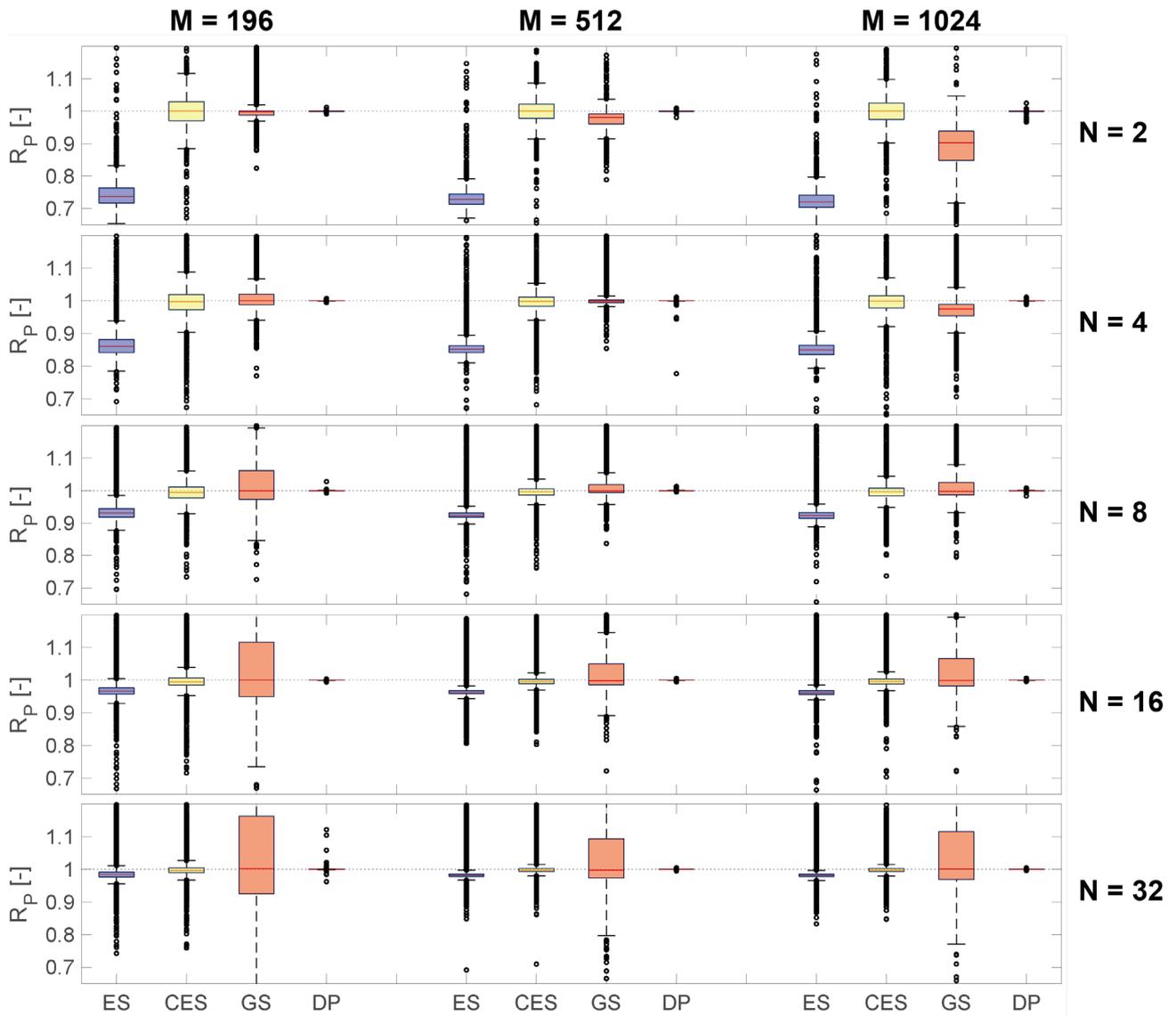

**Fig. 3 Box-and-whisker plot comparing ES (Eigensolver), CES (corrected Eigensolver), GS (GS-PAT), and DP (Diff-PAT) for different combinations of control points (N) and transducer numbers (M).** The box shows the lower quartile, median, and upper quartile of the dataset (the total number of sample size or control points is 1000 × N) for each algorithm, and the maximum whisker length is set as 1.5 times the interquartile range. The black circles indicate outliers (i.e., values greater than the whisker length). Analysed using Matlab ® R2020a (Update 4) with Statistics and Machine Learning Toolbox ® Ver 11.7.

and the median value improves to unity for M = 196, 512 and 1024 when N = 2.

In addition, the performance of Eigensolver and the corrected Eigensolver improves when the control point number increases from N = 2 to N = 32 in all types of arrays. Specifically, the interquartile (IQ) range value for the corrected Eigensolver method when N = 2 and M = 196 is 0.058; however increasing N to 32 improves the IQ range value to 0.015. We attribute this to the variability of the target amplitude in a given control point geometry. As the number of control points increases, the average difference between the minimum and maximum amplitude decreases. This is partially because of how the target amplitude is assigned when control points are randomly generated. The sum of the amplitudes assigned to the control points is always equal to a constant (see the Methods section), and the difference between the minimum and maximum amplitude for a given control point geometry decreases from 523, 283, and 70 Pa for N = 2, 8, and 32, respectively (M = 196 with corrected Eigensolver). This statement is supported by further analysis of the dataset of the corrected Eigensolver when N = 2 and M = 196. When the control point geometries lie between the IQ range, the average amplitude difference between the control points is 443 Pa. However, beyond the IQ range, the average amplitude difference between the control points increases to 603 Pa.

As shown in Fig. 3, the performance of GS-PAT decreases as the number of control points N increases. When N = 2 and M = 196, the IQ range is 0.012; however, when N is increased to 32, the IQ range increases to 0.238. By contrast, increasing the number of transducers to M = 512 improves the performance of GS-PAT, and the IQ range drops to 0.119 when N = 32 and M = 512. These results are consistent with the observations made by Placensia et al. who claimed that GS-PAT is comparable to Eigensolver up to N = 8 when M = 512. However, as the transducer number increases to M = 1024, the performance of GS-PAT drops again (IQ range increases to 0.089 and 0.146 for N = 2 and 32 when M = 1024).

Conventional solvers (i.e. Eigensolver, corrected Eigensolver, and GS-PAT) show a trade-off between the numbers of transducers and control points. In contrast, Diff-PAT outperforms all solvers in all cases evaluated in Fig. 3. Diff-PAT is robust against amplitude variability within the control point geometries and can consistently achieve the target amplitude despite the reduction in transducer numbers. In addition, while the Eigensolver type method and GS-PAT use both amplitude and phase control, Diff-PAT outperforms both methods with only phase control.

**Comparing computational time of each solver**

The computational time of the solver is insignificant for non-active PATs; however, some applications of PATs require consideration of the computational efficiency. Fig. 4 compares the computational time of each solver, and all of the solvers were executed on the same high-end desktop computer (4.2 GHz Core i7-7700K, 64 GB RAM). Eigensolver and corrected Eigensolver (Fig. 4a and b) both show similar trends in computational time, and an increase in control points (N) does not have a significant effect on their computational time. However, doubling the number of transducers results in a magnitude increase in the computational time for both Eigensolver and corrected Eigensolver (e.g. when N = 2, the computational time for Eigensolver is 6.2, 110, and 830 ms for M = 196, 512, and 1024, respectively). GS-PAT achieves the lowest computational time of all solvers in all of the conditions (lowest computational time is 0.12 ms with N = 2 and M=196, and the highest computational time is 21 ms with N=32 and M = 1024). The authors of GS-PAT have reported up to 17000 geometries per second[33], making it well suited for application requiring rapid calculations. However, as demonstrated in Fig. 3 and 4, GS-PAT is inaccurate in comparison to other solvers, and the regions in which GS-PAT can achieve both accuracy and speed is limited.

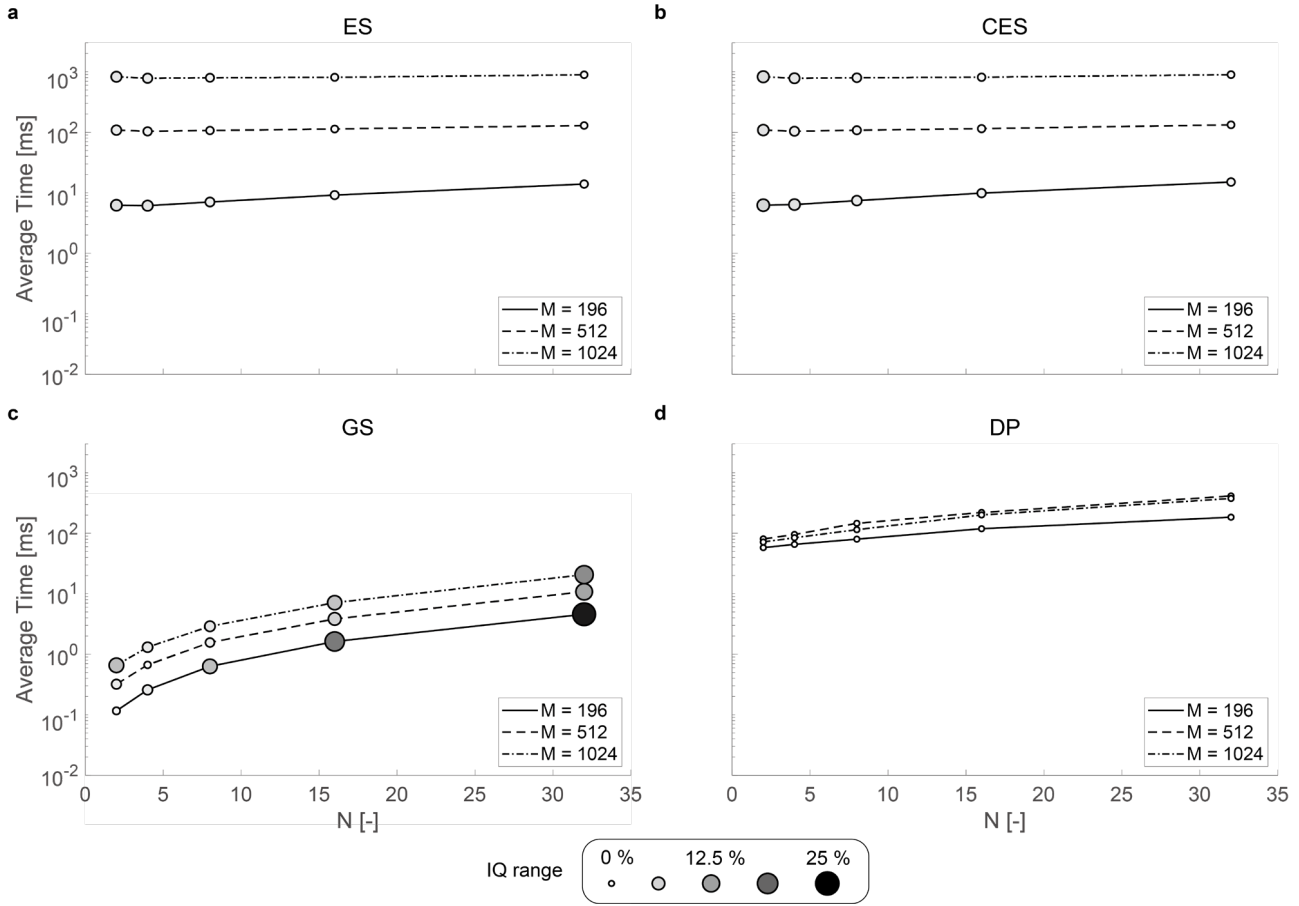

**Fig. 4 Comparing the average computational time for solving one geometry using each solver.** The sample size is 1000 geometries, and solved the same dataset as in Fig. 3. The magnitude of IQ range from Fig. 3 is transposed to the marker size and darkness. White and small marker indicate low IQ range, and larger and darker mark indicate high IQ range. (**a**) Eigensolver (ES), (**b**) corrected Eigensolver (CES), (**c**) GS-PAT (GS) and (**d**) Diff-PAT (DP). (a)-(c) is a C++ code, where (d) was run on Python. The performance was evaluated on the same desktop computer.

Diff-PAT does not have a computational efficiency that is as high as that of GS-PAT; however, its computational time is comparable to the Eigensolver type optimiser with M = 512. When M = 512 and $N \leq 4$, Diff-PAT is faster than the Eigensolver type method and when M = 1024, Diff-PAT is faster than the Eigensolver method in all cases of N. In addition, Diff-PAT scales well with an increasing number of transducers as shown in Fig. 4d. Given that the demand for high transducer number PATs is increasing[18,50–53]: Diff-PAT is already more suited to optimise large transducer number PATs than an Eigensolver type method, both in terms of accuracy and computational efficiency. Furthermore, there are abundant applications of PATs where the accurate reconstruction of the acoustic field is prioritised over computational speed[18,25–27,54].

**Application of Diff-PAT for Acoustic Metamaterials**

In this section, we further explore the versatility of Diff-PAT by employing Diff-PAT for the optimisation of acoustic metamaterial[42–45,55,56]. Specifically, we optimised a type of acoustic metamaterial called a phase plate[42–45] which is used underwater. It has a significantly higher number of elements than PATs[44], and is well-suited for testing the capabilities of Diff-PAT. The IASA developed by Melde et al. is one of the most popular optimisation methods for phase plates. Similar to the GS-PAT and IBP methods, IASA also applies Gerchberg-Saxton algorithm[34]. To create a phase plate version of Diff-PAT, the acoustic pressure field was assumed to be a plane wave[42] and the angular spectrum approach[57] was used to calculate the propagating acoustic field in the loss function (see Methods for details).

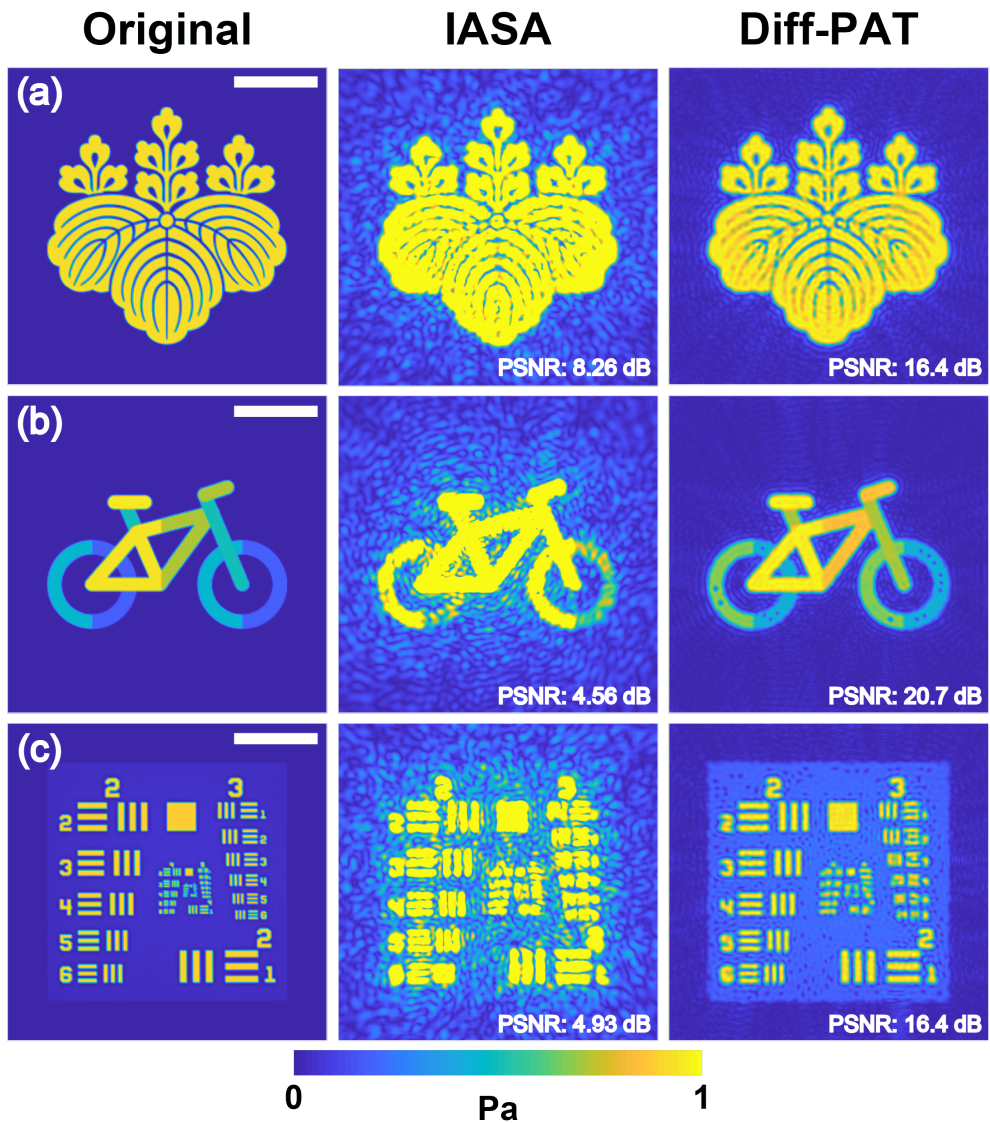

**Fig. 5 Comparison of acoustic holograms optimised using Diff-PAT and IASA.** The resolution of the original image is 256s×256 pixels. An underwater 2 MHz transducer with a 35 mm diameter was assumed, and the pixel size was set to 150 μm. (**a**) Emblem of the University of Tsukuba (**b**) a bicycle (**c**) USAF 1951 resolution test chart. The scale bar shows 10 mm.

Fig. 5 compares the holograms optimised using IASA and Diff-PAT (see Supplementary Information for the acoustic holograms of IASA and Diff-PAT). The reconstruction accuracy of Diff-PAT is clearly higher than that of IASA. A simple visual inspection verifies that the acoustic hologram optimised using IASA, shown in Fig. 5a-c, has many artefacts and does not resolve the test image well (Fig. 5c). In contrast, the acoustic hologram optimised using Diff-PAT has a significantly improved image with minimal artefacts, as shown in Fig. 5a-b and the USAF 1951 resolution test chart is clearly resolved in Fig 5c. Quantitatively, the peak signal-to-noise ratio (PSNR) values for Diff-PAT are 16.4, 20.7 and 16.4 dB (Fig 5a-c, respectively), and they are at least 8 dB higher the respective PSNRs of IASA. In practical applications of phase plates, these optimised acoustic holograms are encoded and fabricated into a plate[42] using a 3D printer. The superior accuracy of Diff-PAT with significantly reduced artefacts can enhance performance of a wide range of acoustic systems used in medicine[1–3], biology[4–8], and engineering[9–11] applications.

In conclusion, we presented Diff-PAT, which optimises only the phase of an acoustic hologram using automatic differentiation. We demonstrated that Diff-PAT achieves superior accuracy in comparison with conventional optimisers in both PATs and metamaterials. In addition, contrary to common belief; the phase only optimiser can achieve better accuracy than optimiser with both amplitude and phase control. For PATs with high transducer numbers (M = 1024), Diff-PAT is more suitable than Eigensolver type methods, both in terms of accuracy and computational efficiency. In future, this improved acoustic hologram optimiser will improve the performance of many applications by providing higher quality acoustic holograms. In addition, this study, along with Peng et al.[33], demonstrated the effectiveness of implementing automatic differentiation outside of machine learning. We hope that these results will further facilitate the use of automatic differentiation in a wide range of fields and applications that still rely on numerical differentiation.

## Methods

**Total Acoustic Pressure Calculation**

The total pressure was evaluated by $p_t = \sum_{m=0}^{M} \frac{P_{ref}}{d(x_c, x_t)} D(\theta) e^{j(kd(x_c, x_t) + \phi_{n,m})}$ where $M$ is the total number of transducers, $P_{ref}$ is the reference pressure amplitude; $d(x_c, x_t)$ is the Euclidean distance between $x_c$ and $x_t$, and $D(\theta) = \frac{2J_1(kr\sin(\theta))}{kr\sin(\theta)}$ is the far-field directivity function for the piston source based on the angle, θ. Moreover, θ is evaluated as the angle between the transducer normal and $x_c$; $r = 5$ mm is the transducer radius, and $J_1$ is the Bessel function of the first kind of order 1. $k = \frac{2\pi f}{c_0}$ is the wavenumber given the acoustic frequency, $f = 40$ kHz and the speed of sound is $c_0$. $P_{ref}$ and $c_0$ are assumed to be 1.98 Pa at 1 m (12 V Pk-Pk[30]) and $346$ ms$^{-1}$, respectively.

**Random generation of control points and amplitude**

Control points, $x_c$ are randomly generated such that $x$, $y$, and $z$ coordinates are within the region of interest (ROI), i.e. $[-0.05, 0.05]$ m from the centre of the array. For single-sided arrays, the z-axis centre is set at $z = 0.1$ m. The physically achievable pressure amplitude was predetermined by making a singular focal point ($\phi = -\frac{2\pi f}{c_0}[d(x_r, x_t) - d(0, x_r)]$) at the vertex of ROI, $x_r$ (8 points in total). The average acoustic pressures around the vertices for single-sided (M = 196), single-axis (M = 512), and large single-sided (M = 1024) PATs are 1512, 3812, and 4121 Pa, respectively. Amplitude $A_c$ for each control point is randomly assigned such that the minimum pressure amplitude is 10 Pa, and the amplitude at each control point sums up to the average acoustic pressure for the single focus of each transducer array.

**Application of Diff-PAT to Metamaterial**

For the optimisation of acoustic metamaterial (phase plate), an underwater transducer with a resonance frequency of 2 MHz, diameter of 35 mm, and input amplitude of 1 Pa was assumed at the surface. The input pressure field was assumed to be a plane wave, and the initial guess of the phase was set to zero for all elements. The speed of sound in water was assumed to be 1480 $\mathrm{ms}^{-1}$, and the pixel size was assumed to be 150 μm. For simplicity, the acoustic transmission loss through the plate was considered negligible, and the angular spectrum was solved using methods described by Zeng & McGough[57] (following the implementation of k-Wave[58]). Both IASA and Diff-PAT were programmed in Python 3.6.9, and the phase plate version of Diff-PAT used Tensorflow[38] (ver. 2.3.0) to differentiate the loss function automatically. We also used the Adam optimiser in TensorFlow with the same hyperparameter setting as in the PAT version. IASA was calculated by following the steps described by Melde et al.[42], using the angular spectrum approach proposed by Zeng & McGough[57]. The propagation distance from the transducer to the image plane was 20 mm, and the iteration number for the optimiser (both IASA and Diff-PAT) was set to 200. The loss function was $\sum_x^N \sum_y^N |A_c(x,y) - |p(x,y)||$ where $A_c(x,y)$ represents the target image with a screen resolution (N) of 256× 256 pixels. The optimised acoustic hologram was exported from Python and the resultant acoustic pressure fields shown in Fig. 5 were calculated using k-Wave (ver. 1.3 *angularSpectrumCW*)[58] on Matlab. Finally, the PSNR was calculated using the Matlab Image Processing Toolbox.

**Data Availability**

The data that support the findings of this study are available within this article. The data for transducer arrangements, control point geometries and amplitude, output phase by Diff-PAT, and the results from all solvers are available in figshare at [URL].

**Code Availability**

The program code for optimisation and analysis is also made available at the link provided in Data Availability.

**References**

1. Jiménez-Gambín, S., Jiménez, N., Benlloch, J. M. & Camarena, F. Holograms to Focus Arbitrary Ultrasonic Fields through the Skull. *Phys. Rev. Appl.* **12**, 1 (2019).

2. Sapozhnikov, O. A., Tsysar, S. A., Khokhlova, V. A. & Kreider, W. Acoustic holography as a metrological tool for characterizing medical ultrasound sources and fields. *J. Acoust. Soc. Am.* **138**, 1515–1532 (2015).

3. Baresch, D. & Garbin, V. Acoustic trapping of microbubbles in complex environments and controlled payload release. *Proc. Natl. Acad. Sci.* **117**, 15490–15496 (2020).

4. Ma, Z. *et al.* Acoustic Holographic Cell Patterning in a Biocompatible Hydrogel. *Adv. Mater.* **32**, 1–6 (2020).

5. Kreider, W. *et al.* Characterization of a multi-element clinical HIFU system using acoustic holography and nonlinear modeling. *IEEE Trans. Ultrason. Ferroelectr. Freq. Control* **60**, 1683–1698 (2013).

6. Baudoin, M. *et al.* Folding a focalized acoustical vortex on a flat holographic transducer: Miniaturized selective acoustical tweezers. *Sci. Adv.* **5**, 1–7 (2019).

7. Baudoin, M. *et al.* Spatially selective manipulation of cells with single-beam acoustical tweezers. *Nat.*

*Commun.* **11**, 1–10 (2020).

8. Baresch, D., Thomas, J.-L. & Marchiano, R. Observation of a single-beam gradient force acoustical trap for elastic particles: acoustical tweezers. *Phys. Rev. Lett.* **116**, 024301 (2016).

9. Kruizinga, P. *et al.* Compressive 3D ultrasound imaging using a single sensor. *Sci. Adv.* **3**, (2017).

10. Melde, K. *et al.* Acoustic Fabrication via the Assembly and Fusion of Particles. *Adv. Mater.* **30**, 1–5 (2018).

11. Gong, Z. & Baudoin, M. Particle assembly with synchronized acoustic tweezers. *Phys. Rev. Appl.* **12**, 1 (2019).

12. Hoshi, T., Takahashi, M., Iwamoto, T. & Shinoda, H. Noncontact tactile display based on radiation pressure of airborne ultrasound. *IEEE Trans. Haptics* **3**, 155–165 (2010).

13. Long, B., Seah, S. A., Carter, T. & Subramanian, S. Rendering volumetric haptic shapes in mid-air using ultrasound. *ACM Trans. Graph.* **33**, 1–10 (2014).

14. Marzo, A. & Drinkwater, B. W. Holographic acoustic tweezers. *Proc. Natl. Acad. Sci.* **116**, 84–89 (2018).

15. Marzo, A. *et al.* Holographic acoustic elements for manipulation of levitated objects. *Nat. Commun.* **6**, 8661 (2015).

16. Ochiai, Y., Hoshi, T. & Rekimoto, J. Three-Dimensional Mid-Air Acoustic Manipulation by Ultrasonic Phased Arrays. *PLoS One* **9**, e97590 (2014).

17. Marzo, A., Caleap, M. & Drinkwater, B. W. Acoustic Virtual Vortices with Tunable Orbital Angular

Momentum for Trapping of Mie Particles. *Phys. Rev. Lett.* **120**, 44301 (2018).

18. Prisbrey, M., Guevara Vasquez, F. & Raeymaekers, B. 3D ultrasound directed self-assembly of high aspect ratio particles: On the relationship between the number of transducers and their spatial arrangement. *Appl. Phys. Lett.* **117**, (2020).

19. Youssefi, O. & Diller, E. Contactless Robotic Micromanipulation in Air Using a Magneto-Acoustic System. *IEEE Robot. Autom. Lett.* **4**, 1580–1586 (2019).

20. Puskar, L. *et al.* Raman acoustic levitation spectroscopy of red blood cells and Plasmodium falciparum trophozoites. *Lab Chip* **7**, 1125 (2007).

21. Hosseinzadeh, V. A., Brugnara, C. & Holt, R. G. Shape oscillations of single blood drops: applications to human blood and sickle cell disease. *Sci. Rep.* **8**, 1–8 (2018).

22. Yurduseven, O., Cooper, K. & Chattopadhyay, G. Point-Spread-Function (PSF) Characterization of a 340-GHz Imaging Radar Using Acoustic Levitation. *IEEE Trans. Terahertz Sci. Technol.* **9**, 20–26 (2019).

23. Hasegawa, K. & Kono, K. Oscillation characteristics of levitated sample in resonant acoustic field. *AIP Adv.* **9**, 035313 (2019).

24. Andrade, M. A. B., Camargo, T. S. A. & Marzo, A. Automatic contactless injection, transportation, merging, and ejection of droplets with a multifocal point acoustic levitator. *Rev. Sci. Instrum.* **89**, 125105 (2018).

25. Ochiai, Y., Hoshi, T. & Rekimoto, J. Pixie dust: graphics generated by levitated and animated objects

in computational acoustic-potential field. *ACM Trans. Graph.* **33**, Article 85 (2014).

26. Morales, R., Marzo, A., Subramanian, S. & Martínez, D. LeviProps: Animating levitated optimized fabric structures using holographic acoustic tweezers. *Proc. 32nd Annu. ACM Symp. User Interface Softw. Technol.* 651–661 (2019) doi:10.1145/3332165.3347882.

27. Sahoo, D. R. *et al.* JOLED: A Mid-air Display based on Electrostatic Rotation of Levitated Janus Objects. in *Proceedings of the 29th Annual Symposium on User Interface Software and Technology - UIST '16* 437–448 (ACM Press, 2016). doi:10.1145/2984511.2984549.

28. Uno, Y. *et al.* Luciola: A Millimeter-Scale Light-Emitting Particle Moving in Mid-Air Based On Acoustic Levitation and Wireless Powering. *Proc. ACM Interactive, Mobile, Wearable Ubiquitous Technol.* **1**, 1–17 (2018).

29. Fushimi, T., Marzo, A., Drinkwater, B. W. & Hill, T. L. Acoustophoretic volumetric displays using a fast-moving levitated particle. *Appl. Phys. Lett.* **115**, 64101 (2019).

30. Hirayama, R., Martinez Plasencia, D., Masuda, N. & Subramanian, S. A volumetric display for visual, tactile and audio presentation using acoustic trapping. *Nature* **575**, 320–323 (2019).

31. Marzo, A. *et al.* Holographic acoustic elements for manipulation of levitated objects. *Nat. Commun.* **6**, 8661 (2015).

32. Marzo, A. & Drinkwater, B. W. Holographic acoustic tweezers. *Proc. Natl. Acad. Sci.* **116**, 84–89 (2018).

33. Plasencia, D. M., Hirayama, R., Montano-Murillo, R. & Subramanian, S. GS-PAT: High-speed Multi-

point Sound-fields for Phased Arrays of Transducers. *ACM Trans. Graph.* **39**, (2020).

34. Gerchberg, R. W. A practical algorithm for the determination of phase from image and diffraction plane pictures. *Optik (Stuttg).* **35**, 237–246 (1972).

35. Sakiyama, E. *et al.* Midair Tactile Reproduction of Real Objects. in *Haptics: Science, Technology, Applications* (eds. Nisky, I., Hartcher-O'Brien, J., Wiertlewski, M. & Smeets, J.) 425–433 (Springer International Publishing, 2020).

36. Peng, Y., Choi, S., Padmanaban, N. & Wetzstein, G. Neural Holography with Camera-in-the-loop Training. *ACM Trans. Graph. (SIGGRAPH Asia)* **39**, 1–14 (2020).

37. Chakravarthula, P., Peng, Y., Kollin, J., Fuchs, H. & Heide, F. Wirtinger holography for near-eye displays. *ACM Trans. Graph.* **38**, (2019).

38. Abadi, M. *et al.* TensorFlow: Large-Scale Machine Learning on Heterogeneous Distributed Systems. (2016).

39. Paszke, A. *et al.* Pytorch: An imperative style, high-performance deep learning library. in *Advances in neural information processing systems* 8026–8037 (2019).

40. Bradbury, J. *et al.* JAX: composable transformations of Python+NumPy programs. *Http://Github.Com/Google/Jax* (2018).

41. Güne¸, A., Baydin, G., Pearlmutter, B. A. & Siskind, J. M. Automatic Differentiation in Machine Learning: a Survey. *J. Mach. Learn. Res.* **18**, 1–43 (2018).

42. Melde, K., Mark, A. G., Qiu, T. & Fischer, P. Holograms for acoustics. *Nature* **537**, 518–522 (2016).

43. Melde, K. *et al.* Acoustic Fabrication via the Assembly and Fusion of Particles. *Adv. Mater.* **1704507**, 1704507 (2017).

44. Cox, L., Melde, K., Croxford, A., Fischer, P. & Drinkwater, B. W. Acoustic Hologram Enhanced Phased Arrays for Ultrasonic Particle Manipulation. *Phys. Rev. Appl.* **12**, 64055 (2019).

45. Franklin, A., Marzo, A., Malkin, R. & Drinkwater, B. W. Three-dimensional ultrasonic trapping of micro-particles in water with a simple and compact two-element transducer. *Appl. Phys. Lett.* **111**, (2017).

46. Kingma, D. P. & Ba, J. L. Adam: A method for stochastic optimization. *3rd Int. Conf. Learn. Represent. ICLR 2015 - Conf. Track Proc.* 1–15 (2015).

47. Fushimi, T., Hill, T. L., Marzo, A. & Drinkwater, B. W. Nonlinear trapping stiffness of mid-air single-axis acoustic levitators. *Appl. Phys. Lett.* **113**, 034102 (2018).

48. Prisbrey, M. & Raeymaekers, B. Ultrasound Noncontact Particle Manipulation of Three-dimensional Dynamic User-specified Patterns of Particles in Air. *Phys. Rev. Appl.* **10**, 034066 (2018).

49. Morris, R. H., Dye, E. R., Docker, P. & Newton, M. I. Beyond the Langevin horn: Transducer arrays for the acoustic levitation of liquid drops. *Phys. Fluids* **31**, 101301 (2019).

50. Inoue, S. *et al.* Acoustical boundary hologram for macroscopic rigid-body levitation. *J. Acoust. Soc. Am.* **145**, 328–337 (2019).

51. Hasegawa, K., Yuki, H. & Shinoda, H. Curved acceleration path of ultrasound-driven air flow. *J. Appl. Phys.* **125**, 1–7 (2019).

52. Ochiai, Y., Hoshi, T. & Rekimoto, J. Three-Dimensional Mid-Air Acoustic Manipulation by Ultrasonic

Phased Arrays. *PLoS One* **9**, e97590 (2014).

53. Hasegawa, K., Qiu, L., Noda, A., Inoue, S. & Shinoda, H. Electronically steerable ultrasound-driven long narrow air stream. *Appl. Phys. Lett.* **111**, (2017).

54. Prisbrey, M. & Raeymaekers, B. Aligning High-Aspect-Ratio Particles in User-Specified Orientations with Ultrasound-Directed Self-Assembly. *Phys. Rev. Appl.* **12**, 1 (2019).

55. Polychronopoulos, S. & Memoli, G. Acoustic levitation with optimized reflective metamaterials. *Sci. Rep.* **10**, 4254 (2020).

56. Memoli, G. *et al.* Metamaterial bricks and quantization of meta-surfaces. *Nat. Commun.* **8**, 1–8 (2017).

57. Zeng, X. & McGough, R. J. Evaluation of the angular spectrum approach for simulations of near-field pressures. *J. Acoust. Soc. Am.* **123**, 68–76 (2008).

58. Treeby, B. E. & Cox, B. T. k-Wave: MATLAB toolbox for the simulation and reconstruction of photoacoustic wave   fields. *J. Biomed. Opt.* **15**, 21314 (2010).

**Acknowledgements**

This work was funded by Pixie Dust Technologies, Inc. We would like to acknowledge the authors of GS-PAT for making the code of the conventional optimisation algorithm available for comparison on C++. We would also like to acknowledge the developers of k-Wave. We would like to thank Tatsuya Minagawa for assisting with the visualization of the PATs in Fig. 1. Fig. 5b was designed by Freepik from Flaticon; we have license to modify and use the image for commercial or non-commercial purposes. We would like to thank Editage (www.editage.com) for English language editing.

**Author contributions**

T.F., K.Y., and Y.O. designed the study; T.F. and K.Y. performed the research; T.F. and K.Y. analysed the data; and T.F., K.Y., and Y.O. wrote the paper.

**Competing interests**

A patent application (JP2020-167367A) was filed in relation to this publication. Yoichi Ochiai has multiple unpaid advisory position for governmental/non-governmental bodies in Japan. This work was funded by Pixie Dust Technologies, Inc.

**Materials & Correspondence**

Correspondence and requests for material should be addressed to T.F. (email: tfushimi@slis.tsukuba.ac.jp).

**Figure Legends**

**Fig. 1 System overview for Diff-PAT.** Loss function is evaluated by comparing the target and current acoustic amplitude, and automatic differentiation is used to calculate the derivative of loss function.

**Fig. 2 Array configuration and the convergence plot.** (**a**) Configuration of single-sided and single-axis PATs. (**b**) Convergence of optimisation function $\mathcal{L}$ by the iteration number, control point number (N) and transducer number (M). The convergence is evaluated by $R_p$. The data point and the error bar show the mean and standard deviation of $R_p$, respectively, for the population of randomised 1000 sets of control points.

**Fig. 3 Box-and-whisker plot comparing ES (Eigensolver), CES (corrected Eigensolver), GS (GS-PAT), and DP (Diff-PAT) for different combinations of control points (N) and transducer numbers (M).** The box shows the lower quartile, median, and upper quartile of the dataset (the total number of sample size or control points is 1000 × N) for each algorithm, and the maximum whisker length is set as 1.5 times the interquartile range. The black circles indicate outliers (i.e., values greater than the whisker length). Analysed using Matlab ® R2020a (Update 4) with Statistics and Machine Learning Toolbox ® Ver 11.7.

**Fig. 4 Comparing the average computational time for solving one geometry using each solver.** The sample size is 1000 geometries, and solved the same dataset as in Fig. 3. The magnitude of IQ range from Fig. 3 is transposed to the marker size and darkness. White and small marker indicate low IQ range, and larger and darker mark indicate high IQ range. (**a**) Eigensolver (ES), (**b**) corrected Eigensolver (CES), (**c**) GS-PAT (GS) and (**d**) Diff-PAT (DP). (a)-(c) is a C++ code, where (d) was run on Python. The performance was evaluated on the same desktop computer.

**Fig. 5 Comparison of acoustic holograms optimised using Diff-PAT and IASA.** The resolution of the original image is 256×256 pixels. An underwater 2 MHz transducer with a 35 mm diameter was assumed, and the pixel size was set to 150 μm. (**a**) Emblem of the University of Tsukuba, (**b**) a bicycle, and (**c**) the USAF 1951 resolution test chart. The scale bar shows 10 mm.